# Priority-aware Gray-box Placement of Virtual Machines in Cloud Platforms

**[1]Xia Liu, [2]Li Fan**

[*1]Beijing University of Posts and Telecommunications, Computer Science School, Beijing, China
[2]Beijing University of Posts and Telecommunications, Computer Science School, Beijing, China

Email: liuxia06882@gmail.com

**Abstract** –Virtual machine (VM) placement is very important for cloud platforms. While techniques, such as live virtual machine migration, are very useful to balance the load in the data centers, they are expensive operations. In this position paper, we propose to minimize the chance of the load hot spots in the data center by applying the workload patterns of the VMs in the virtual machine placement algorithms - place VMs that require a lot of same type of resource across different physical servers. In this way, the resource competition of VMs on the same physical server is significantly mitigated. Meanwhile, we also consider the priorities of applications and VMs in our virtual machine placement algorithms.

## 1. Introduction

Virtualization technology that decouples the application execution environment from the underlying hardware is the enabling technique for Cloud Computing, which is growing rapidly and expects to be worth $176.8 billion by 2015 according to Gartner [14]. By multiplexing the hardware resources among many applications hosted in virtual machines (VMs), it can achieve higher resource utilization and thus reduce the infrastructure deployment cost. However, virtualization technology alone is far from enough for the Cloud Computing to be successful. Due to the dynamic application demand, the load in the data center are often imbalanced, giving rise to unnecessary resource deficits on the overloaded physical servers, even though the overall resources in the data centers are enough to handle the demand of all the hosted applications. For example, application instances running on the same physical server may sometimes require much more resources than the server can provide, causing severe performance degradation and application QoS violation, while at other times the application instances receive so few requests that the resources of the underlying physical machines are totally wasted. Thus, another crucial technology for Cloud Computing is the dynamic resource management – allocating resources to applications based on need.

Many techniques have been proposed to allocate more resources to applications when needed. For example, live VM migration was invented to move the overloading VMs to physical servers with more available resources and minimize the down time [2, 18]. Also, ghost VMs that are pre-deployed and cached in the memory were proposed to increase the number of the service replicas quickly enough to absorb the surging application demand [11, 20]. Although useful and necessary, all these techniques incur heavy-weight operations. For example, migration consumes high CPU and network bandwidth. Ghost VMs have the same memory cost as working VMs, though CPU and network cost are trivial.

In order to reduce the opportunity of applying these expensive techniques, in this position paper, we explore how to minimize the chance of load hot spots in the data centers in the first place. We propose a new virtual machine placement strategy which tries to put the application instances (We assume that each VM only hosts one application instance for good isolation among application instances. Thus, referring to an application instance is equivalent to referring to an VM.) that potentially require a lot of resources of the same type on different physical machines. The observation behind is that by putting the application instances that require different type of resources in the same physical server, the chance that this physical server gets overloaded is much lower since the competition among the collocated VMs on that server is mitigated.

We can analyze the application server logs to obtain the precise application workload patterns. However, in most environments, cloud providers are not allowed to access the log statistics in the guest. In order to address this issue, we propose the "gray-box" VM placement strategy, which requires only the high level workload characteristics, such as whether the application is CPU-intensive workload, or network-intensive, in stead of the detailed workload pattern. These high-level workload characteristics can be acquired through dynamic monitoring or application profiling when deploying the applications as discussed in Section II-A.

Meanwhile, we also take into account of the priorities of different applications when deciding the placement of virtual machines. The quality of service (QoS) requirements vary for different applications. Furthermore, different components of an application may have different QoS requirements. Modern applications oftentimes consist of multiple components, each hosted in a separate VM. When placing the virtual machines, we need to make sure that VMs with higher priorities are less interrupted by the load hot spots in the data centers.

In summary, in this position paper, we take the high-level workload patterns of the VMs and different application priorities into consideration, and propose heuristic algorithms for virtual machine placement to reduce the chance of load hot spots in the data centers. In the remaining of the paper, we first discuss the virtual machine placement problem under this environment and show that it is NP-hard. Then we explain our heuristic algorithms to solve the problem. Finally we give the conclusion and introduce the future work.

## 2. Problem Statement

### 2.1. High-level Workload Pattern

In virtualized cloud computing, a physical server is multiplexed with many VMs to enhance the utilization of the resources and reduce the cost. This paradigm, at the same time, also increases the opportunity of overloading of the underlying servers due to the dynamic demand of the VMs. For example, according to the work [17], there are many transient and some long-term memory overcommitment in production data centers. An ideal optimization is to place VMs whose workload variations are evenly distributed temporally on the same physical server so that the overall resource requirements of the hosted VMs can still be satisfied. However, the unpredictability of the demand makes it extremely challenging (if not impossible). Furthermore, precise and detailed workload pattern is hard to acquire in current cloud computing environments since usually cloud providers are not allowed to access the application logs in the guest OS for security and legal reasons.

However, we believe that it is still possible to optimize the virtual machine placement without the detailed and precise workload patterns of VMs. Our experience with practical application systems indicates that applications tend to require heavy resource capacity only for certain type of resources. For example, relatively less CPU capacity is needed for streaming applications that usually ask for a lot of network bandwidth and for database applications that often require many disk I/O operations. Also, data analysis applications often require much more CPU capacity than other type of resources. Thus, even though the workload for each type of resources is still dynamic, actually only the demand of certain type of resources will change dramatically and thus matter a lot in the decision-making. We refer the type of resources that are needed heavily by the applications as the dominant resources, and this workload information as high-level workload pattern of applications. Note that, it is possible that some applications have more than one dominant resource.

We argue that the high-level workload pattern can be utilized to improve the virtual machines placement in the data centers. For example, we can place VMs whose dominant resources are different on the same hosts so that each VM is potentially be able to consume large portion of the resources of the host regardless of other collocated VMs (since they demand more of other type of resources). Moreover, it is feasible and relatively easy to obtain these high-level workload patterns in the cloud environment. For IaaS cloud platforms, cloud providers can monitor the resource consumption of the VMs and obtain the high-level workload pattern. For PaaS cloud, it is also possible to acquire this information from the customers by asking simple questions of the deployed applications. For private cloud, it is even possible to get the precise and detailed workload information since cloud providers are usually the owners of the hosted applications. Since we only require very limited workload information of the VMs, we refer it as "graybox" virtual machine placement, which is discussed in details in Section 3.

### 2.2. VM Priorities

Cloud providers also need to meet the quality of service requirements of the hosted applications. Some applications, such as the banking or stock trading systems, are very sensitive to performance degradation and have higher QoS requirements, while some other applications are more tolerable to resource shortage. We must give different priorities to VMs of different applications. Meanwhile, modern applications usually consist of many components, each running in separate VMs. Thus, for the same applications, the priorities of the VMs may be different.

The priorities of VMs usually depend on how important the applications are to the application providers and how much money they are willing to pay to ensure the QoS satisfaction. There can be multiple mechanisms to assign the priorities to the VMs. In this position paper, we assume that the priorities are already assigned and are expressed as non-positive number ranging from 1 to 100; the bigger the numbers are, the higher the priorities the VMs have.

### 2.2. Modeling

Let $M$ be the number of physical servers, $A$ the number of applications and $V$ the number of virtual machines in the data centers. Each VM $v_i$ has a dominant resource set $DR(v_i)$, and is associated with priority $PR(v_i)$. We consider four type of resources: CPU, memory, network and disk, each denoted by $r$, $r = 1; 2; 3; 4$ respectively. Each VM $v_i$ has resource demands of different types, expressed as vector $<_1 d(v_i);$

$_2 d(v_i);$

$_3 d(v_i);$

$_4 d(v_i) >.$ Let
$_r u(p_i)$ and

r
c(p$_i$) denote the utilization and capacity of physical server p$_i$ for each resource type r respectively. We need to find a placement the matrix P in which, each item P$_{i,j}$=1 if the VM v$_i$ is placed at the physical server p$_j$ ; P$_{i,j}$=0 otherwise. Let

r__

u be the average utilization of the physical servers for resource type r in the data center. The utilization of physical server p$_j$ for resource type r can be calculated as:

## 3. Methodology

The placement of virtual machines in the data centers usually includes two part: initial placement and incremental placement. Initial placement intends to place a set of VMs onto the physical servers in the data centers to balance the load of different type of resources. Due to the potentially large number of VMs and physical servers in the data centers (e.g., mega data centers), it is very important to find an efficient algorithm for the VM placement. Incremental placement is usually employed to mitigate the load hot spots detected in the data centers. In this section, we explore both the initial placement and incremental algorithms for the placement of virtual machines in the data centers. We explain our methodology to solve the virtual machine placement problem considering both the high-level workload pattern and priorities of VMs discussed above.

### 3.1 Initial Placement

VMs are categorized into five sets, S$_r$, r = 1, 2, 3, 4, 5. Each set S$_r$ (r = 1, 2, 3,4) contains VMs that are CPU-intensive, memory-intensive, network-intensive, or disk-intensive, according to value of r respectively. Thus, VMs in the same set are likely to compete intensively for the corresponding resource. The set S$_5$ contains VMs that do not have obvious preference of any type of resources. Note that, it is possible that some VMs can appear at multiple sets, indicating that they have multiple dominant resources, though we assume this is not typical.

In the initial placement algorithm, we try to distribute the VMs from the same set S$_r$ (r = 1, 2, 3, 4) across the different physical servers at the best effort. In this way, the VMs on the same physical servers would have different type of dominant resources and thus are less likely to cause the overloading on the physical servers. Meanwhile, in order to better satisfy the QoS requirements of VMs with higher priorities, we try to place the high-priority VMs on physical servers with more capacity available.

```
1.   For r = 1 to 5, put the VMs into set S_r if the resource r is the
     dominant resource of the VM.
2.   Sort the VMs decreasingly in the set S_r by their priorities
3.   If r < 5
4.       Sort the physical servers decreasingly by the capacity of
         resource r
5.   End If
6.   While S_r ≠ ∅
7.       For j = 1 to M
8.           Pick up the unplaced VM v_i from the list of set S_r
9.           If demand of VM v_i can be satisfied by the physical
             server p_j
10.              Place the VM v_i at the physical server p_j .
11.              Mark that the VM v_i has been placed.
12.              Remove the VM v_i from set S_r.
13.              Resort the physical servers decreasingly by the
                 capacity of resource r
14.              If S_r = ∅.
15.                  r++;
16.                  Goto 1.
17.              Else
18.                  j++;
19.              End If
20.          Else
21.              j++;
22.              Goto 7.
23.          End If
24.      End For
25.  End While
26.  End For
```

Figure 1. Initial placement algorithm

Figure 1 shows the pseudo code of the initial placement algorithm. The physical servers are sorted according to their capacity of the type of the resource in each loop. For the VMs of the same set, it picks up the VM with highest priority that has not been placed yet and try to put it on a physical server that has the largest capacity available of the dominant resources of the VM. Note that, even though the physical servers are sorted by their capacity of the dominant resource of the VM, the selected physical server also needs to satisfy the demand of the VMs for other type of resources. Thus the first physical server is not necessarily always the best candidate and more physical servers need to be checked sometimes. Every time a VM is placed, the list of the physical servers are resorted to make sure that physical servers with high capacity available are placed first. It scans the list of physical servers repeatedly until all the VMs in the set are placed on some physical server. Thus, each physical server would have similar number of VMs from the same set S$_r$, r = 1, 2, 3, 4, 5.

In the initial placement algorithm, the VMs that are in the set S$_5$ are placed regardless of their resource preferences since they do not have dominant resources. In order to feed the VMs with higher priority better, more optimization can be done, such as picking up the physical servers that have more capacity of all type of resources needed first, etc., which is not shown in Figure 1. Another issue is that, for VMs that have more than one dominant resource, they would be placed according to the first dominant resource only since it is marked "placed" in the loop of the first dominant resource. In order to handle this issue, we can also do some optimization for these kind of VMs, like placing them based on the type of dominant resource with the largest significance, etc., which is not shown in Figure 1 either. Or if some VMs have too many dominant resources, we can simply put them in the set S$_5$.

## 3.2 Incremental Placement

The incremental placement algorithm also considers the high-level workload pattern and the priorities of the virtual machines. When a physical server is overloaded, we need to migrate some virtual machines to other physical servers with enough capacity available. The main problem for incremental placement is, which VMs to be migrated and to which physical servers should these VMs be migrated?

Migration of VMs incurs performance degradation in the transition of the starting and stopping of VMs [2]. Thus, it is preferable to migrate the VMs that have lower priorities on the overloading physical servers. As for where to migrate the VMs, we can also select physical servers that have more capacity of the resource because of which the physical server is overloaded. The algorithm for the incremental placement is shown in Figure 2.

When picking up the VMs to be migrated, the VM with the larger demand of the resource due to which physical server $p_j$ is overloaded is preferred if there are multiple VMs with the same priority. In this way, the resource contention on $p_j$ can be mitigated faster.

## 4. RELATED WORK

The virtual machine placement problem is very important in the cloud computing systems and many algorithms have been developed before for this problem. Some works, like the [20, 11, 18], propose heuristic algorithms to find the physical servers with more capacity to place or migrate VMs that require more resources incrementally. In our incremental algorithm, we consider the priorities of the VMs as well as the capacity of physical servers. Moreover, we also propose the initial virtual machine placement algorithm.

Many algorithms [10, 12, 13, 9, 6, 4, 7] are proposed to place and migrate the virtual machines in the data centers that consider the relationship between the VMs and focus on reduce the network distance and bandwidth among the VMs. For example, [10] proposes a virtual machine placement and migration approach to minimize the data transfer time consumption in the data centers. [12] considers the inter-VM dependencies and the underlying network topology into VM migration decisions in their VM migration algorithm. Authors in [13] introduce a virtual machine placement algorithm to decrease the communication distance between VMs, improving the energy-efficient and scalability of data centers. [9] proposes network-aware VM placement algorithm based on the network architecture and traffic pattern to address the scalability problem in the data centers. Virtual machine placement algorithm in [6] addresses the scalability problem in the data centers taking into account of the traffics among the VMs. The work in [4] considers the VM migration in a bandwidth oversubscribed tree network to balance the load. In [7], authors proposed placement algorithm to solve the problem for modern data centers spanning placement of application computation and data among available server and storage resources. Our work differentiate from them mainly in that we consider all type of resources in the data center, and try to place the VMs that are with similar workload at different physical servers to reduce the chance of overloading of servers. In addition, we also consider the priorities among different applications and virtual machines.

1. For each overloaded physical server $p_j$, order its VMs increasingly by their priorities on the overloaded
2. While the $p_j$ is overloaded
3.   Pick up the VM $v_i$ from the list.
4.   Find a physical server with the largest capacity of dominant resource of $v_i$ that can satisfy the total resource requirements
5.   Migrate the VM $v_i$ to that physical server.
6.   Update the status of that physical server and $p_j$.
7. End While

Figure 2. Incremental placement algorithm.

Some other works focus on the energy-related issue when dealing with the virtual machine placement problem in the data centers. In [19], a VM placement approach is introduced to minimize the total resource wastage, power consumption and thermal dissipation costs. [5] addresses the energy efficient VM placement problem in cloud architecture with multidimensional resources. The work of [3] tries to lower the power consumption while fulfilling performance requirements we propose a flexible and energy-aware framework for the allocation of virtual machines in a data center. Authors in [16] propose an innovative application-centric energy-aware strategy for virtual machine allocation. The proposed strategy ensures high resource utilization and energy efficiency through VM consolidation while satisfying application QoS. Finally, in [8] a new algorithm Dynamic Round-Robin (DRR), is proposed for energy-aware virtual machine scheduling and consolidation. Our work does not address the energy issue in the data centers. However, it would be very interesting to consider the energy-related issues in our future work.

Finally, [21] develops an algorithm to minimize VM migrations in over-committed data centers. While our work also try to reduce the opportunity of overloading in the data center so that less migrations operations are needed to balance the load in the data center, we achieve this by making use of the high-level workload pattern of the VMs and applications. Also, this work does not consider the priorities of the VMs and applications. The work in [15] is similar with ours in that it also tries to utilize the workload pattern to place the VMs in the data center. However, they rely on more precise and detailed workload pattern which is only possible for private cloud platforms, while in our work, only high-level workload pattern is needed. Also, we consider the priority of the VMs while they do not.

## 5. Conclusion and Future work

In this position paper, we propose to utilize the high-level workload pattern to place virtual machines in data centers to reduce the chance of the hot spots in the data centers. We develop heuristic algorithms in which virtual machines with the same dominant resources are placed at different physical servers at the best effort so that the resource competition among the VMs on a physical server is mitigated. Our heuristic algorithms cover both initial VM placement and incremental placement. Meanwhile, our algorithms also consider the priorities of the VMs and try to place VMs with higher priorities to physical servers with more capacity of the dominant resources of these VMs.

We are implementing our algorithms and build our system for extensive evaluation, which is the main task of the future work. Meanwhile, we would also consider the energy-related problem in the future.